# Updating the Routing Position in Adhoc Networks by using Adaptive Position

Ravi Kumar Poluru[1]    T. Sunil Kumar Reddy[2]    D.Nagaraju[3]

[1](CSE, P.G Scholar, Sir Vishveshwariah institute of science and technology, India)

[2](CSE, Associate Professor, Sir Vishveshwariah institute of science and technology, India)

[3] (CSE, Assistant Professor, Sir Vishveshwariah institute of science and technology, India)

**Abstract:** *In geographic routing, nodes to theirinstantaneous neighbors call for to maintain up-to-date positions for making successfulforwarding decisions. The geographic location coordinates of the nodes by the periodic broadcasting of beacon packetsis atrendy method used by the majority geographic routing protocols to preserve neighbor positions. We contend and display that periodicbeaconing apart from of the node mobility and traffic patterns in the network is not nice-looking from both update cost and routingrecital points of view. We recommend the Adaptive Position Update (APU) strategy for geographic routing, which enthusiasticallyadjusts the occurrence of position updates based on the mobility dynamics of the nodes and the forwarding patterns in the system. APUis based on two easy principles: 1) nodes whose arrangements are harder to guess update their positions more recurrently (and viceversa), and (ii) nodes faster to forwarding paths update their positions more recurrently (and vice versa). Our speculative analysis, whichis validated by NS2 simulations of a well-known geographic routing procedure, Greedy Perimeter Stateless Routing Protocol (GPSR),shows that APU can drastically reduce the update cost and pick up the routing performance in terms of packet delivery ratio andregular end-to-end delay in assessment with periodic beaconing and other recently proposed updating schemes. The benefits of APUare additional confirmed by responsibility evaluations in realistic network scenarios, which account for localization error, practical radiopropagation, and sparse system.*

## 1. INTRODUCTION

The routing protocols are increasing their popularity in the field of positioning device that is interested to use in mobile adhoc networks. The principle behind the protocol is that it has to select the node from the list of nodes, which is closer to the destination. Here the forwarding decision is taken by the local knowledge. By taking this characteristics, location based routing protocol are highly useful and healthy to frequent updates in the networks. Pertaining to this the forwarding decision is taken on the go, each node select the next node which has the finest possibility to taken as a next node. Based on several studies these routing protocols offer good performance improvements over topology-based routing protocols such as AODV [1] and DSR [2].

In geographical routing the transmission strategy require following information:

1. The position of the final destination to receive the packet.
2. The neighboring node position.

The grid location system which gives the information about previous node by querying it. To get the next node location each node exchanges its information with neighboring nodes. This will help the each node to build a route map.

However, in some situations where nodes are moving or on and off, the local topology rarely remains unchanged. Hence it is important to the nodes to broadcast their location information to all its neighbors. This type transformation we called as beacons. In most geographical routing protocols these beacons are helpful to maintaining the neighboring list at each node.

## 2. LITERATURE SURVEY

In 1973 the U.S. Defense academy began the DARPA packet radio network project. This project maintains the communication routes in a network using radios. This PRNET [4] routing protocol uses a form of distance vector routing every node broadcasts its information for every 7.5 seconds. The packet header contains the source and destination address information, the number of nodes visited so far from the source, and the remaining nodes which have to be visited to reach the destination. Based on the header information nodes update their routing tables. The link protocol uses hop by hop acknowledgements or passive acknowledgements from received packets.

The problem in the distance vector routing is it forms routing loops. In order to eliminate this problem, Perkins and Bhagwat have proposed adding sequence numbers to the routing updates in Destination–Sequenced Distance Vector Routing protocol. The age of the information in the routing table can be compared by using sequence numbers, and allow each nodes to select the fresh information.

## 3. ADAPTIVE POSITION UPDATE

The assumptions which are made in our work are:

1. In the network all nodes are conscious about their own location and velocity.
2. In the network the communication would be in bidirectional.
3. The current location and velocity of the nodes should be included in the beacon.
4. Data packets can piggyback position.

Upon initializing, each node has to transmit its beacon information about their location and velocity. The neighboring nodes which receive the information can be stored; the local topology of each node can continuously update. Only those nodes from the neighbor list are treated as possible nodes for data forwarding. Thus, the beacons play a major part in maintaining a precise representation of the local topology. APU [3] deploys two beacon triggering mutual exclusive rules which are discussed in the following.

### 3.1 Predicating the mobile rule:

The mobile rule adapts the frequency to beacon generation rate in which nodes can change their motion. This characteristic of motion which are included in the broadcast to neighbor nodes. The nodes have to update their status by changing their frequency. On the other side nodes moving slowly cannot update their status. In our scheme, upon receiving a beacon update from a node N, each of its neighbors records node N's current position and velocity and periodically track node N's location using a simple prediction scheme based on linear kinematics (discussed below). Based on this position estimate, the neighbors can check whether node N is still within their transmission range and update their neighbor list accordingly. The goal of the MP rule is to send the next beacon update from node N when the error between the predicted location in the neighbors of N and node N's actual location is greater than an acceptable threshold.

We use a simple location prediction scheme based on the physics of motion to estimate a node's current location. Note that, in our discussion, we assume that the nodes are located in a 2D coordinate system with the location indicated by the N and M coordinates. However, this scheme can be easily extended to a 3D coordinate system. Table 1 illustrates the notations used in the rest of this discussion.

TABLE 1: NOTATIONS FOR MOBILITY PREDICTION

| Variables | Definitions |
|---|---|
| $N_p, M_p$ | The coordinate of node p at time T |
| $U_a, U_b$ | The velocity of node p at time T |
| $Z_T$ | The time of the last beacon packet |
| $Z_C$ | The current time |
| $N_q, N_q$ | The prediction time of node q at time T |

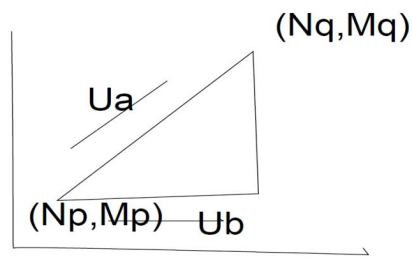

Fig1: mobility prediction example

As shown in Fig. 1, given the position of node i and its velocity along the N and M axes at time Zl, its neighbors can estimate the current position of x, by using the following equations.

$$N^x_p = N^x_p + (Z_d - Z_T) * U^X_a \quad (1)$$

$$M^x_p = M^x_p + (Z_d - Z_T) * U^x_b \quad (2)$$

### 3.2 Learning the on-Demand Rule

The MP rule exclusively may not be sufficient for maintaining an accurate local topology. Consider the example illustrated in Fig. 2, where node A moves from P1 to P2 at a constant velocity. Now, assume that node A has just sent a beacon while at P1. Since node B did not receive this packet, it is unaware of the existence of node A. Further, assume that the AER is sufficiently large such that when node A moves from P1 to P2, the MP rule is never triggered. However, as seen in Fig. 2 node A is within the communication range of B for a significant portion of its motion. Even then, neither A nor B will be aware of each other. The MP rule is not sufficient to maintain local topology. Take an example which is given in fig 2, where node A motion from P1 to P2. Now assume that node A sent beacon when it is at P1. Node B doesn't aware of this transmission and existence of node A. later In fig 2 node B is within the communication range of node A. Now, in situations where neither of these nodes is transmitting data packets, this is perfectly fine since they are not within communicating range once A reaches P2. However, if either A or B was transmitting data packets, then their local topology will not be updated and they will exclude each other while selecting the next hop node. In the worst case, assuming no other nodes were in the vicinity, the data packets would not be transmitted at all.

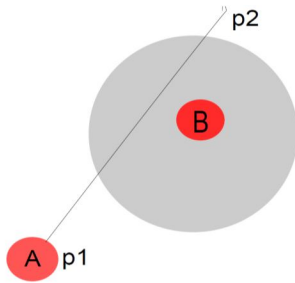

Fig 2: An example illustrating a drawback of the MP rule.

Fig. 3a demonstrates the network topology before node A starts send data to node P. The solid lines in the figure denote that both ends of the link are aware of each other. The initial possible routing path from A to P is A-B-P. Now, when source A sends data packets to B, both C and D receive the data packet from A. As A is a new neighbor of C and D, according to the ODL rule, both C and D will send back beacons to A. As a result, the links AC and AD will be discovered. Further, based on the location of the destination and their current locations, C and D discover that the destination P is within their one-hop neighborhood. Similarly, when B forwards the data packet to P, the links BC and BD are discovered. Fig. 3b reflects the enriched topology along the routing path from A to P.

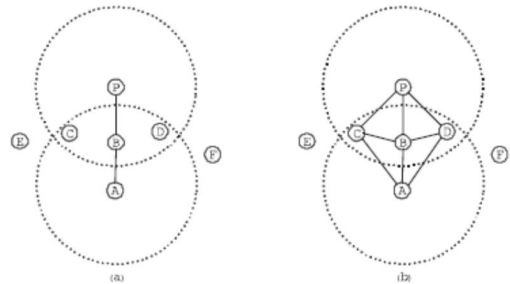

Fig 3: an example for ODL rule

Note that, though E and F receive the beacons from C and D, respectively, neither of them responds back with a beacon. Since E and F do not lie on the forwarding path, it is futile for them to send beacon updates in response to the broadcasts from C and D. In essence, ODL aims at improving the accuracy of topology along the routing path from the source to the destination, for each traffic flow within the network.

### 3.3 Updating the Adaptive Position

In this section, we analyze the performance of the proposed beaconing strategy, APU [5][6]. We focus on two key performance measures: 1) update cost and 2) local topology accuracy. The former is measured as the total number of beacon broadcast packets transmitted in the network. The latter is collectively measured by the following two metrics:

**.3.3.1 Unknown neighbor ratio.** This is defined as the ratio of the new neighbors a node is not aware of, but that are within the radio range of the node to the total number of neighbors.

**3.3.2 False neighbor ratio**. This is defined as the ratio of obsolete neighbors that are in the neighbor list of a node, but have already moved out of the node's radio range to the total number of neighbors. The unknown neighbors of a node are the new neighbors that have moved in to the radio range of this node but have not yet been discovered and are hence absent from the node's neighbor table. Consider the example in Fig. 4, which illustrates the local topology of a node X at two consecutive time instants. Observe that nodes A and B are not within the radio range R of node X at time t. However, in the next time instant (i.e., after a certain period t), both these nodes have moved into the radio range of X. If these nodes do not transmit any beacons, then node X will be unaware of their existence. Hence, nodes A and B are examples of unknown neighbors.

On the other hand, false neighbors of a node are the neighbors that exist in the node's neighbor table but have actually moved out from the node's radio range (i.e., these nodes are no longer reachable). Consider the same example in Fig. 4. Nodes C and D are legitimate neighbors of node X at time t. However, both these nodes have moved out of the radio range of node X in the next time instant. But, node X would still list both nodes in its neighbor table. Consequently, nodes C and D are examples of false neighbors.

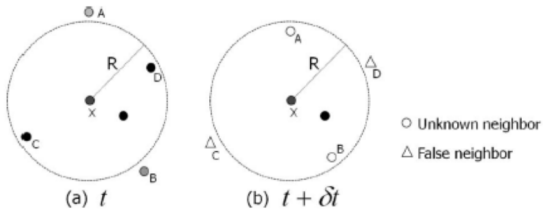

Fig 4: example for unknown and false neighbors.

Note that, the existence of both unknown and false neighbors adversely impacts the performance of the geographic routing protocol. Unknown neighbors are ignored by a node when it makes the forwarding decision. This may lead to suboptimal routing decisions, for example, when one of the unknown neighbors is located closer to the destination than the chosen next-hop node. If a false neighbor is chosen as the next hop node, the transmitting node will repeatedly retransmit the packet without success, before realizing that the chosen node is unreachable (in 802.11 MAC, the transmitter retransmits several times before signaling a failure). Eventually, an alternate node would be chosen, but the retransmission attempts waste bandwidth and increase the delay.

### 3.3 Beacon Overhead

Recall that the two rules employed in APU are mutually exclusive. Thus, the beacons generated due to each rule can be summed up to obtain the total beacon overhead. Let the beacons triggered by the MP rule and the ODL rule over the network operating period be represented by OMP and OODL, respectively. The total beacon overhead of APU, OAPU, is given by

$$O_{APU} = O_{MP} + O_{ODL} \quad (3)$$

Next, we proceed to separately analyze OMP and OODL.

### 3.4 Beacon Overhead Due to the ODL Rule ($O_{ODL}$)

According to the ODL rule, whenever a node overhears a data transmission from a new neighbor, it broadcasts a beacon as a response. In other words, beacons are transmitted in response to data forwarding activities. Let $\chi$ denote the total number of data packet forwarding operations that occur over the network operating period and let $\gamma$ be the average number of beacons that are triggered by each forwarding operation. Now, the total beacons triggered by the ODL rule, $O_{ODL}$, can be represented by

$$O_{ODL} = \chi \cdot \gamma. \quad (4)$$

Next, we proceed to derive $\chi$ and $\gamma$.

**3.4.1 $\chi$ Analysis**. The total number of data packet forwarding operations can be represented as the product of the number of packets generated in the network and the number of times each packet is forwarded. The number of packets generated in the network during a finite time period of $\Gamma$ can be expressed as $\lambda M \Gamma$, where $\lambda$ is the packet generation rate (packets per second) at each source, M is the number of communication pairs (i.e., source-destination pairs). Let H be the average number of hops along the forwarding paths between the source and destination nodes. In other words, each packet is forwarded on average, H times, as it progresses from the source to the destination. Hence, $\chi$ can be represented as

$$\chi = \lambda M \Gamma \cdot H \quad (5)$$

Since $\lambda$, M, and $\Gamma$ are known network parameters, we only need to derive H

Since, the nodes are uniformly distributed in the network (a property of the RDM model [11]), the distance between a source-destination pair is equivalent to the distance between two randomly selected points. In [10], Bettstetter et al. have analyzed the distance between two randomly select points, and formulated the average distance (D) as

$$D = \frac{1}{15}\left[\frac{A^3}{B^2} + \frac{b^3}{B^2} + \sqrt{A^2 + B^2\left(3 - \frac{A^2}{B^2} - \frac{B^2}{A^2}\right)}\right]$$
$$+ \frac{1}{6}\left[\frac{B^2}{A} arcos\left(\frac{\sqrt{A^2 + B^2}}{B}\right)\right.$$
$$\left. + \frac{A^2}{B} arccos\left(\frac{\sqrt{A^2 + B^2}}{A}\right)\right]$$

(6)

where A*B denotes the network dimensions. Based on work [11], given the Euclidean distance D between the source and destination node, the average number of hops between these nodes can be represented as follows:

$$H = \frac{D}{R.\left[1 - \int_0^1 1 - \exp\left(\rho R(arccos(t) - t\sqrt{1 - t^2})\right) dt\right]}$$

(7)

Where ρ is the average node density, which is given by A. B=N.

Combining (5), (6), and (7), we obtain the total number of data packet forwarding operations χ.

### 3.4.2. γ Analysis.

Let δ(t) be the probability that a neighboring node moves out the radio range of a node during a small interval t. In other words δ(t) denotes the link breakage probability [7]. Given that a node has an average of $\rho\pi R^2$ neighbors [8], the number of neighbors that move out of the radio range of anode during the time 1/ λ follows:

Next, we derive δ(t). Intuitively, δ(t) is a function of the mobility pattern [10] of the nodes. The faster the nodes move, the higher is the link breakage probability.

## 4. SIMULATION RESULTS

In this section, we present a comprehensive simulation-based evaluation of APU using the popular NS-2 simulator. We compare the performance of APU with other beaconing schemes. These include PB and two other recently proposed adaptive beaconing schemes in : (i) Distance-based Beaconing [9] and (ii) Speed-based Beaconing (see Section 2).

Table 2
Energy consumption in each operation

| operation | µW.sec/byte[8] | µW.sec |
|---|---|---|
| Point to point send | 0.48*size | +431 |
| Broadcast send | 2.1*size | +272 |
| Point to point recv | 0.12*size | +316 |
| Broadcast recv | 0.26*size | +50 |
| Promiscuous recv | 0.12*size | +83 |
| Promiscuous discard | 0.11*size | +54 |

(The point to point communication uses data rate of 11 mbps. The broadcasting uses data rate of 2 mbps. Therefore, broadcasting costs more energy than point to point sending)

We conduct three sets of experiments. In the first set of simulations, we demonstrate that APU can effectively adapt the beacon transmissions to the node mobility dynamics and traffic load. In addition, we also evaluate the validity of the analytical results derived in Section 4, by comparing the same with the results from the simulations. In the second set of experiments, we consider the impact of real-world factors such as localization errors, realistic radio propagation,and sparse density of the network on the performance of APU. In the third set of experiments, we evaluate the impact of parameter AER (which is from MP component) on the overall performance of APU. This enables us to investigate which component (MP or ODL) contributes to the performance more significantly.

We use two sets of metrics for the evaluations. The first set includes the metrics used in our analysis, viz., beacon overhead and local topology accuracy (false and unknown neighbor ratio), which directly reflect the performance achieved by the beaconing scheme. Note that the beaconing strategies are an integral part of geographic routing protocols. The second set of metrics seek to evaluate the impact of the beaconing strategy on the routing performance. These include: 1) packet delivery ratio, which is measured as the ratio of the packets delivered to thedestinations to those generated by all senders, 2) average end-to-end delay incurred by the data packets, and 3) energy consumption, which measures the total energy consumed in the network.

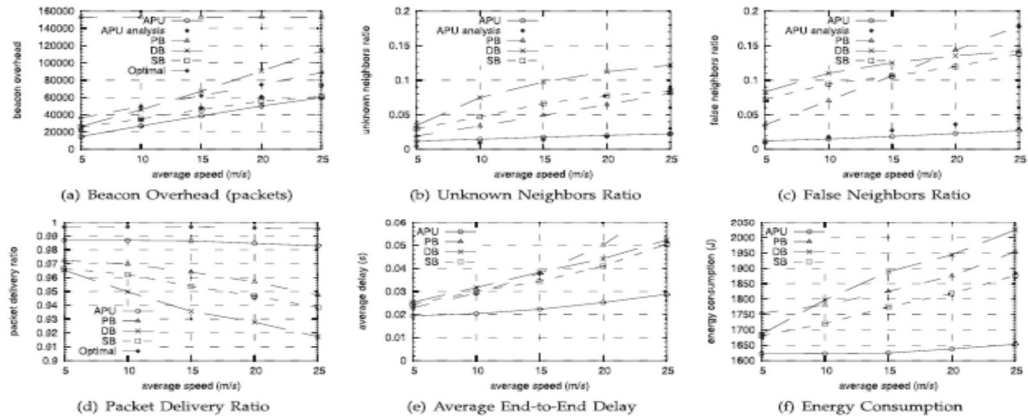

Fig. 5. Impact of node speed on the performance of beaconing schemes.

## 5. CONCLUSIONS

In this paper we have dealt with beacon update policy which is employed in geographical routing protocols to the traffic load and mobility dynamics.

We suggested the adaptive position update strategy to address these problems. Two mutually exclusive rules will be employed by the APU scheme. To estimate the accuracy of the location MP rule adapts the beacon update policy.The nodes along the data forwarding path to maintain an accurate view of the local topology it uses ODL rule by exchanging beacons in response to data packets that are overheard from new neighbors.Weanalyzed the local topology and beacon overhead accuracy of APU and validated thesimulation results with the analytical model. We have embedded APU within GPSR and have compared it with other related beaconing strategies using extensive NS-2 simulations for varying node speeds and traffic load.

## 6. REFERENCES

[1] J. Hightower and G. Borriello, "Location Systems for UbiquitousComputing," Computer, vol. 34, no. 8, pp. 57-66, Aug. 2001.

[2] B. Karp and H.T. Kung, "GPSR: Greedy Perimeter StatelessRouting for Wireless Networks," Proc. ACM MobiCom, pp. 243-
254, Aug. 2000.

[3] L. Blazevic, S. Giordano, and J.-Y.LeBoudec, "A Location BasedRouting Method for Mobile Ad Hoc Networks," IEEE Trans.Mobile Computing, vol. 4, no. 2, pp. 97-110, Mar. 2005.

[4] Y. Ko and N.H. Vaidya, "Location-Aided Routing (LAR) in MobileAd Hoc Networks," ACM/Baltzer Wireless Networks, vol. 6, no. 4,pp. 307-321, Sept. 2002.

[5] T. Camp, J. Boleng, B. Williams, L. Wilcox, and W. Navidi,"Performance Comparison of Two Location Based RoutingProtocols for Ad Hoc Networks," Proc. IEEE INFOCOM,pp. 1678-1687, June 2002.

[6] D. Johnson, Y. Hu, and D. Maltz, The Dynamic Source RoutingProtocol (DSR) for Mobile Ad Hoc Networks for IPv4, IETF RFC 4728,vol. 15, pp. 153-181, Feb. 2007.

[7] C. Perkins, E. Belding-Royer, and S. Das, Ad Hoc On-DemandDistance Vector (AODV) Routing, IETF RFC 3561, July 2003.

[8] J. Li, J. Jannotti, D.S.J.D. Couto, D.R. Karger, and R. Morris, "A\ Scalable Location Service for Geographic Ad Hoc Routing," Proc.ACM MobiCom, pp. 120-130, Aug. 2000.

[9] Z.J. Haas and B. Liang, "Ad Hoc Mobility Management withUniform Quorum Systems," IEEE/ACM Trans. Networking, vol. 7,no. 2, pp. 228-240, Apr. 1999.

[10] A. Rao, S. Ratnasamy, C. Papadimitriou, S. Shenker, and I. Stoica,
"Geographic Routing without Location Information," Proc. ACMMobiCom, pp. 96-108, Sept. 2003.

[11] D.arunkumarreddy and T.Sunilkumarreddy, " AStudy of Asynchronus Routing Protocols in Peer to Peer Networks," International Journal of P2P Networks trends and Technologies (IJPTT) Vol 5 Feb 2014.

[12] T.Sunil Kumar Reddy and D.Nagaraju," Security issues in Dynamic Topological Peer to Peer Networks",International Journal of Communication and Engineering Vol 4 Issue 3 Sep 2013.

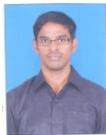Ravi Kumar Poluru is an PG Scholor in the Department of Computer science & engineering, Sir Vishveshwariah Institute of Science and Technology, Madanapalli.  He received the B.Tech degree in Computer Science & Engineering from JNTU University in 2006.His research include mobile computing, wireless networks.

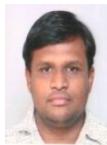T. Sunil Kumar Reddy is an associate professor in the Department of Computer science & engineering, Sir Vishveshwariah Institute of Science and Technology, Madanapalli.  He received the B.Tech degree in Information Technology from Satyabhama University in 2005, the M.Tech degree in Information Technology from V.I.T University in 2007and he is pursuing PhD degree in computer science & engineering from JNTUA University, Anantapur. His research interests include Cloud Computing, High performance computers, Wireless Networks.

D. Nagaraju is an assistant professor in the Department of Computer science & engineering, Sir Vishveshwariah Institute of Science and Technology, Madanapalli.  He received the B.Tech degree in Computer Science & Engineering from JNTU University in 2009, the M.Tech degree in Computer Science  from JNTU Anantapur in 2012. His research include Network Security, Cloud computing.